\documentclass[12pt]{article}
\usepackage{pdproc,epsfig}

  \def\be{\begin{equation}}
  \def\ee{\end{equation}}

\begin{document}

%%%%%%%%%%%%%%%%%

\renewcommand{\thefootnote}{\alph{footnote}}

\title{NEUTRINO MASSES AND MIXING: LEPTONS VERSUS QUARKS
\footnote{Talk given at  
the III International Workshop on: NO-VE "Neutrinos in Venice" 
Fifty years after the Neutrino discovery, 
7-10 Feb 2006, Venice, Italy.}}

\author{ALEXEI YU. SMIRNOV}

\address{Physik-Department T30d, Technische Universit\"at M\"unchen,
James-Franck-Strasse,\\ 
D-85748 Garching, Germany\\
International Centre for Theoretical Physics,  
Strada Costiera 11, 34014 Trieste, Italy\\
Institute for Nuclear Research, RAS, Moscow, Russia\\
 {\rm E-mail: smirnov@ictp.trieste.it, smirnov@ph.tum.de}}

\abstract{
Comparison of properties of
quark and leptons as well as understanding their similarities and 
differences is one of the milestones  
on the way to underlying physics. 
Several  observations, if not accidental,  
can strongly affect the implications: 
(i) nearly tri-bimaximal character of  lepton mixing,
(ii) special neutrino symmetries, (iii) the QLC-relations.
We consider possible connections between quarks and leptons which include  
the quark-lepton symmetry and unification, approximate universality,  
and quark-lepton complementarity.
Presence of new neutrino states and their mixing with
the left or/and  right handed neutrinos can be
the origin of additional differences of quarks and leptons.}

\normalsize\baselineskip=15pt
%%%%%%%%%%%%%%%%%%%%%%%%%%%%%%%%%%%%%%%%%%%%%%%%%%%%%%%%%%%%%%%%%%%%%%%

%%%%%%%%%%%%%%%%%%%%%%%%%%%%%%%%%%%%%%%%%%%%%%%%%%%%%%%%%%%
\section{Introduction}
%%%%%%%%%%%%%%%%%%%%%%%%%%%%%%%%%%%%%%%%%%%%%%%%%%%%%%%%%%%%%%%%%%%%%%%

%%
%The main focus of presentations at this workshop is on  
%determination of the neutrino parameters - masses, mixing angles, 
%CP-violating phases. In this connection, I would like to make one, 
%probably trivial, comment. 
%Clearly, the determination of parameters  is not the ultimate goal. 
%We want to perform  measurements of masses, mixings, phases 
%not just because we want to know certain numbers.  
%We want to know these numbers to understand eventually 
%the underlying physics. From this perspective it is 
%even good that we do not know yet 
%all the masses, mixings, phases:  
%still there is a room and time for predictions and their tests 
%which is the only way to convince that our theory is correct.
%%

One of the key issues on the way to underlying physics
is a comparison of properties of quarks and leptons and  
understanding their similarities and differences.
This comparison has two aspects of the fundamental 
importance:

\begin{itemize}

\item
understanding the fermion masses and mixings;  

\item
uncovering the path  of further unification - unification of  quarks and 
leptons,  particles and forces. 

\end{itemize}

%The unification can be realized
%in context of Grand Unified theories or immediately
%string theories.

Are quarks and leptons similar or fundamentally 
different? Still whole spectrum of possibilities exists 
from the weakly broken quark-lepton universality to 
existence of different  structures and  symmetries 
in these two sectors.

In this paper we confront properties of quarks and leptons.
We then discuss their possible connections:

- symmetry and unification; 

- universality; 

- complementarity; 

- diversity, that is,  existence of new structures
which can produce difference in the two sectors.

%sectors and establishing certain relations between them 
%may give some insight.

\section{Leptons versus quarks}
%%%%%%%%%%%%%%%%%%%%%%%%%%%%%%%%%%%%%%%%%%%%%%%%%%%%%%%%

\subsection{Confronting mixing and masses}
%%%%%%%%%%%%%%%%%%%%%%%%%%%%%%%%%%%%%%%%%%%%%%

To compare mixings in the quark and lepton sector we use the 
standard parametrization of mixing matrices: 
\be
V_{f} = V_{23} (\theta_{23})I_{\delta} V_{13}(\theta_{13}) 
V_{12}(\theta_{12}), ~~~~ f = CKM, ~~PMNS,   
\label{param}
\ee
where $V_{ij}$ is the matrix of rotations in the $ij$- plane, and 
$I_{\delta}$ is the diagonal matrix of the CP-violating phases.
(Notice that $V_{PMNS}$ corresponds to $V_{CKM}^{\dagger}$).

The Table I presents  the mixing angles
in the quark and lepton sectors from the analysis of ref. \cite{bari}. 
Similar results have been obtained by other groups~\cite{sno,sv}. 
Shown are also the sums of the corresponding angles. 
Apparently, the mixing patterns in these two sectors
are strongly different.
The only common  feature is that the 1-3 mixings
(between the ``remote'' generations) are small in both
cases. 

%%%%%%%%%%%%%%%%%%%%%%%%%%%%%%%%%%%%%%%%%%%%%%%%%%%%%%%%%%%%%%%%%%%
\begin{table}
\begin{center}
\begin{tabular}[t]{|c|c|c|c|}
\hline
{\rule[-3mm]{0mm}{12mm}\bf angles} & {\bf quarks} & {\bf leptons} 
{\rule[-4mm]{0mm}{8mm}} & {\bf sum }  \\
\hline
{\rule[-5mm]{0mm}{14mm}\bf $\theta_{12}$} & 
$12.8^{\circ}$ & $33.9^{\circ}$ & $46.7^{\circ} \pm 2.4^{\circ}$ \\
\hline
{\rule[-5mm]{0mm}{14mm}\bf $\theta_{23}$} & 
$2.3^{\circ}$ & $41.6^{\circ}$ & $43.9^{\circ}~^{+ 5.1^{\circ}}_{-3.6^{\circ}}$ \\
\hline
{\rule[-5mm]{0mm}{14mm}\bf $\theta_{13}$} & 
$0.5^{\circ}$ & $< 8.0^{\circ}$ & $< 8.5^{\circ}$ \\
\hline
\end{tabular}
\caption{The best fit  values of mixing angles in the quark and lepton sectors at $m_Z$ scale in degrees.  
Shown are also the sums of the angles with $1\sigma$ error bars. 
}
\end{center}
\end{table}
%%%%%%%%%%%%%%%%%%%%%%%%%%%%%%%%%%%%%%%%%%%%%%%%%%%%%%%%%%%%%%%%%%

Several comments  are in order. 

The b.f. value of the 1-2 mixing angle, 
%from the SNO~\cite{sno} analysis equals
$\theta_{12} = 33.9^{\circ}$,  deviates from the 
maximal mixing by more than $6\sigma$ \cite{sno}.

The 2-3 mixing is consistent  with  maximal one.
A small shift of $\theta_{23}$ from $45^{\circ}$ 
is related to the  excess of 
e-like atmospheric neutrino events in the sub-GeV range detected by 
SuperKamiokande (SK) \cite{atm}. 
It has been found when effects of 1-2 sector were included 
in the analysis \cite{orl}.  
According to \cite{concha}
$\sin^2\theta_{23} = 0.47$ and slightly larger
shift, $\sin^2\theta_{23} = 0.44$,  follows from the analysis \cite{bari}. 
The deviation of the b.f. value from maximal mixing 
is characterized by    
\be
D_{23} \equiv 0.5 - \sin^2\theta_{23} = 0.03 - 0.06. 
\ee

Still large deviation  is allowed:
$- 0.17 < D_{23} < 0.21$ and relative shift can be as large as 
\be
D_{23}/\sin^2\theta_{23} \sim 0.4 ~~~(2\sigma). 
\ee

The 1-3 leptonic mixing is consistent with zero. 
The most conservative $3\sigma$ bound is $\sin^2 \theta_{13} < 0.048$
\cite{bari}, and at $1\sigma$ we have $\sin \theta_{13} < 0.13$.  
The 1-3 mixing is small in a sense that 
\be
\sin \theta_{13} \ll  \sin \theta_{12} \sin \theta_{23} \approx 0.37. 
\ee
So,  apparently the quark feature 
$\theta_{13} \sim \theta_{12} \times \theta_{23}$
does not work here.
Another interesting benchmarks is the ratio of the
solar and atmospheric neutrino mass scales, 
\be
\sin \theta_{13} = \sqrt{r} \equiv 
\sqrt{\frac{\Delta m_{21}^2}{\Delta m_{31}^2}} = 
0.17,   
\ee
which  is allowed at about $2\sigma$ level. An 
additional (model dependent) factor of the order 0.3 - 2   
may appear in this relation. 
Much smaller values of $\sin \theta_{13}$ would
imply most probably certain symmetry of the mass matrix.\\

Let us consider now the masses. 

The latest analysis of the cosmological data 
(including the WMAP 3 years result) gives the upper bound 
on the sum of masses of active neutrinos~\cite{Uros06}  
\be
\sum_i m_i < 0.14~{\rm eV},~~~ 95\%  {\rm C.L}. 
\label{cosmbound}
\ee
which already starts to disfavor the degenerate spectrum of neutrinos. 
 
%%%%%%%%%ffff3%%%%%%%%%%%%%%%%%%%%%%%%%%%%%%%%%%%%%%%%%%%%%%%%%%%%%%%%%%%
\begin{figure}
\vspace*{13pt}
%\leftline{\hfill\vbox{\hrule width 7cm height0.001pt}\hfill}
\begin{center}
\mbox{\epsfig{figure=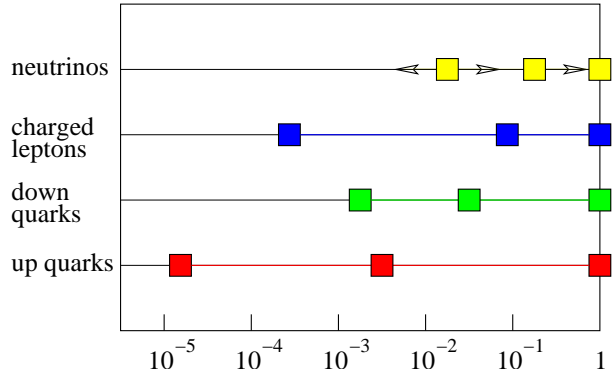,width=8.0cm}}
\end{center}
%\vspace*{1.4truein}             %ORIGINAL SIZE=1.6TRUEIN x 100% - 0.2TRUEIN
%\leftline{\hfill\vbox{\hrule width 5cm height0.001pt}\hfill}
\caption{Mass hierarchies of quarks and leptons.
The mass of heaviest fermion of a given type is taken to be 1.}
\label{ratios}
\end{figure}
%%%%%%%%%%%%%%%%%%%%%%%%%%%%%%%%%%%%%%%%%%%%%%%%%%%%%%%%%%%%%%%%%%%

On the other hand, if the Heidelberg-Moscow result~\cite{hm} is confirmed 
and if it is due to exchange of the light Majorana neutrinos,  the 
neutrino mass spectrum should be strongly degenerate with 
a common mass $m_0 \sim (0.2 - 0.6)$ eV. This would be in conflict with 
the bound (\ref{cosmbound}).

The solar and the atmospheric mass differences squared 
give the lower bound on ratio of the second and third neutrino masses: 
\be
\frac{m_2}{m_3} \geq \sqrt{r} 
= 0.15 - 0.20.
\label{ratiom}
\ee
This should be compared with ratios for charged leptons and quarks (at $m_Z$
scale):
$m_\mu/m_\tau =  0.06$, $m_s/m_b = 0.02 - 0.03$,  $m_c/m_t = 0.005$.
Apparently, the neutrino hierarchy (\ref{ratiom})  is the weakest one.
This is consistent with possible mass-mixing relation: 
large mixings are associated to weak mass  hierarchy.

In fig.~\ref{ratios} we show the mass ratios for three generations.
The strongest hierarchy and geometric relation $m_u \times m_t \sim m_c^2$
exist for the upper quarks.  It seems the observed pattern  
of masses is an  interplay of some regularities (flavor alignment) and randomness 
(``anarchy'').
That may indicate the perturbative picture when 
the lowest order masses and  mixing are universal,  whereas corrections 
have more complicated (``random'') flavor structure. 

In what follows we will discuss certain observed features 
which can strongly affect 
interpretation of the results. 

%%%%%%%%%%%%%%%%%%%%%%%%%%%%%%%%%%%%%%%%%%%%%%%%%%%%%%%%%%
\subsection{Tri-bimaximal mixing} 
%%%%%%%%%%%%%%%%%%%%%%%%%%%%%%%%%%%%%%%%%%%%%%%%%%%%%%%%%%%%

Experimental results are in a very good agreement with the so called 
tri-bimaximal mixing \cite{tbm}. 
The corresponding mixing matrix is 
\be
U_{tbm} = U_{23}^m U_{12}(\theta_{12}) =  
\frac{1}{\sqrt{6}}
\left(\begin{array}{ccc}
2 & \sqrt{2} & 0\\
-1 & \sqrt{2} & \sqrt{3}\\
 1 & - \sqrt{2} & \sqrt{3}
\end{array}
\right), 
\label{tbimax}
\ee
where $\sin^2 \theta_{12} = 1/3$ is about $1\sigma$ larger than  the best   
experimental fit value.    
Here $\nu_2$ is tri-maximally mixed: 
in the middle column three flavors mix maximally,
whereas  $\nu_3$ (third column) is bi-maximally mixed.
Mixing parameters turn out to be some simple
numbers $0,~ 1/\sqrt{3},~ 1/\sqrt{2}$ and can appear as Clebsh-Gordan 
coefficients. 

In the case of normal mass hierarchy ($m_1 \approx 0$)
the mass matrix which leads to the tri-bimaximal mixing 
has the following form 
\begin{equation}
m_{\nu} \approx 
\frac{m_3}{2} 
\left(\begin{array}{ccc}
0 & 0 & 0\\
0 & 1 & - 1\\
0 & - 1 & 1
\end{array}
\right) +
%\frac{\sqrt{\Delta m_{21}^2}}{3}
\frac{m_2}{3}
\left(\begin{array}{ccc}
1 & 1 & 1\\
1 & 1 & 1\\
1 & 1 & 1
\end{array}
\right), 
\label{tribi}
\end{equation}
where $m_2 \approx \sqrt{\Delta m_{21}^2}$ and $m_3 \approx \sqrt{\Delta 
m_{31}^2}$. It is the sum of two singular matrices 
with certain symmetries. The later gives a  hint of its origin. 

Matrix (\ref{tbimax}) was motivated by certain geometric consideration. 
If description of the data by (\ref{tbimax}) is not accidental 
and certain principle/symmetry is behind, we should conclude on  
substantial differences in the quark and lepton sectors.  
Though some models have been  constructed which reproduce the 
tri-bimaximal mixing and  include also quark~\cite{model}. 

\subsection{Complementarity}
%%%%%%%%%%%%%%%%%%%%%%%%%%%%%%%%%%%%%%%%%%%%%%%%%%%%%%%

According to the Table I, the sums of the mixing  angles of quarks 
and leptons for the 1-2 and 2-3 generations agree with $45^{\circ}$. 
The quark and lepton mixings sum up to maximal mixing \cite{qlc,qlc1}. 
Possible implications of this result called  
the quark-lepton complementarity relation (QLC) 
will be considered in sect. 3.3. 
Notice that the QLC relations written for angles are
are essentially parametrization independent. Indeed, 
due to smallness of 1-3 mixings in the quark and lepton sectors 
the relations can be written as 
$\arcsin(V_{us}) + \arcsin(V_{e2}) = \pi/4$. 
The mixing matrix elements $V_{us}$ and  $V_{e2}$ are  
physical parameters.

\subsection{Neutrino symmetry}
%%%%%%%%%%%%%%%%%%%%%%%%%%%%%%%%%%%%%%%%%%%%%%%%%%%%%%%%%

Several observations may testify for special symmetry(ies)
associated to neutrinos. In particular, 
\begin{itemize}

\item
maximal or nearly maximal 2-3 mixing, 

\item
zero 1-3 mixing,  

\end{itemize} 
both indicate toward the same underlying symmetry. 
Both features  
can be consequences of the $\nu_{\mu} - \nu_{\tau}$ permutation symmetry of the 
neutrino mass matrix \cite{mutau} in the flavor basis.
The permutation symmetry can be  a part of, {\it e.g.},  discrete $S_3$,  
$A_4$ or $D_4$ groups which in turn,  are the subgroups of continuous 
SO(3).

Important fact is that the symmetry is realized for neutrinos only, and
only in the flavor basis where the charge lepton
mass matrix is diagonal. The symmetry is 
broken in the charged lepton sector by 
inequality of masses of muon and tau lepton.
Realization of this symmetry in specific gauge models
faces some generic problems. 
Model should be constructed in such a way that  the symmetry
is weakly broken in the neutrino sector but
strongly broken for the charged leptons. 
This implies  different transformation properties of the
right handed components of neutrinos and charged
leptons, since the left components form the SU(2) doublets. 
This, in turn,  contradicts the L-R symmetry,   
and consequently, the $SO(10)$ type of unification. 
Still such symmetry transformations can be consistent with the
$SU(5)$  unification. Alternatively, one can consider
more sophisticated fermionic or/and Higgs sectors.

It is also non-trivial to extend the symmetry
to the quark sector which  prevents from any simple Grand Unification.
A modification of the
$\nu_\mu \leftrightarrow \nu_\tau$ symmetry
has been proposed recently that
can be  the universal symmetry of quarks
and leptons \cite{anj05}.
The symmetry is formulated in the basis which
differs from the flavor basis and therefore
should be considered as the $2-3$ family symmetry.
It is argued that beside maximal (large) 2-3 leptonic
mixing, smallnes of the
$V_{cb}$ element of the CKM-mixing matrix testifies
for this symmetry as well.  

The 2-3 symmetry implies the following universal
form of the mass matrices:
\be
M = 
\left(\begin{array}{ccc}
X & A & A\\
A & B &  C\\
A & C & B
\end{array}
\right) + \delta m, 
\label{muniv}
\ee
where small corrections, $\delta m \ll B$,  are  of the same order for
leptons and quarks.

The 2-3 symmetry does not contradict
mass hierarchy which depends on particular
values of parameters in the matrix (\ref{muniv}).
To get the hierarchical mass spectrum of the charged
fermions (quarks and leptons) one should take
\be
B_{q,l} \approx C_{q,l}, ~~~~ X_{q,l} \ll A_{q,l}
\ll B_{q,l}.
\label{cond1}
\ee
The corresponding matrices are diagonalized by
nearly maximal 2-3 rotation. The physical
CKM mixing is small
(zero in the limit $\delta m \rightarrow 0$).
Large lepton mixing
requires small 2-3 rotation from the neutrino mass matrix.
This can be achieved if
\be
C_{\nu} \ll B_{\nu}, ~~~
C_{\nu} < |\delta m_{22} - \delta m_{33}|.
\label{cond2}
\ee
Furthermore, correct neutrino mass split can be obtained 
if   $X_{\nu} \approx B$, and 
neutrinos have quasi-degenerate spectrum. 
So, essentially the mass matrices of neutrinos and
charged fermions are strongly different; moreover,
large lepton mixing is not the consequence of the
2-3 symmetry but result of tuning of paremeters
of the zero order matrix and corrections. 
Apparently additional symmetries/principles should be
introduced to explain properties
(\ref{cond1}, \ref{cond2}).\\

Generic feature is that
introduction of symmetry is motivated by  maximal
or nearly maximal lepton mixing. However
realizations of the symmetry in a majority of
gauge models show that
large mixing appears eventually as a result of tuning
of parameters and {\it not as consequence of symmetry}.
%Result does not correspond to expectation. 
This clearly makes whole context to be inconsistent.

Two remarks are in order.  

(i) Symmetry is realized in terms of the
mass (Yukawa coupling) matrices.
It turns out that  structure of the mass matrix is very sensitive to
even small deviations of the 2-3 mixing from
maximal and 1-3 mixing from zero.
Taking the best fit values of parameters from \cite{bari}
$\sin^2 \theta_{13} = 0.01$, $\sin^2 \theta_{23} = 0.44$, 
we obtain the matrix of the absolute values of masses in meV \cite{renata}:
\be
M =
\left(\begin{array}{ccc}
3.2 & 6.0 & 0.6\\
... & 24.8  &  21.4\\
... & ... & 30.7
\end{array}
\right) 
\label{mnum}
\ee
which should be compared with the symmetry matrix
(\ref{muniv}). Notice that in contrast to (\ref{muniv}) the 12 and 13 elements
are strongly different and 33- element is greater than  22 element  by
$20-25 \% $.

(ii) The present measurements
admit substantial deviations of $\theta_{23}$ from
maximal and $\theta_{13}$ from zero.
That, in turn,  allows even stronger deviation
of the matrix from the symmetric form.

So, it is not excluded that neutrino symmetry approach is simply misleading.

\subsection{Additional structure?}
%%%%%%%%%%%%%%%%%%%%%%%%%%%%%%%%%%%%%%%%%%%%%%%%%%%%%%%%

The features discussed above: tri-bimaximal mixing, 
neutrino symmetry, quark-lepton complementarity 
may indicate that quarks and leptons are 
fundamentally different and some additional structures exist 
that  lead to this difference. 

The main question here is whether
these features/relations are real or accidental?
``Real'' in a sense that simple and direct  symmetry 
or principle exist which lead to the relations. 
``Accidental'' in a sense  that relations are an interplay (sum)
of several independents effects or contributions.

Quarks and leptons have similar gauge structure,
which establishes clear correspondence of the
leptons and quarks. On the other hand, the quarks
and leptons have strongly different
mass and mixing patterns. 

The hope is that all particular features
of neutrino mass spectrum  and lepton mixing
can be reduced eventually to the neutrality of neutrinos: zero
electric and color charges.
This neutrality opens unique possibility for neutrinos 
to 

-  have the Majorana mass terms, and

-  mix with singlets of the SM symmetry group.

Both features are realized in the seesaw mechanism~\cite{sees}.
As we will see,  the second one
may have two different effects:
(i) modify the mass matrix of active neutrinos,
(ii) produce certain  dynamical effects 
on the neutrino conversion
(if new states are light).  

Is this enough to explain all 
salient properties of neutrinos?
Do the data really indicate existence of new
physics structure
(new particles, interactions, symmetries)? 
Is this additional structure the seesaw, 
or something beyond seesaw is involved?

In this connection a general context could
be that beyond the SM apart from the RH neutrinos
some other fermions (singlets of the SM symmetry group)
exist. These fermions can have various origins
in physics beyond the SM, being related  to
Grand Unification, supersymmetry, existence of extra dimension,
{\it etc.}. Existence of large number of singlets 
is a generic consequence of string theory.  
Masses of these singlets
can be essentially at any scale, from zero to the Planck  mass.
They can mix in general with both  LH and
RH neutrino components.

The singlets and their mixing  with SM neutrinos
may be a missed structure which
explains the difference of quark and lepton properties on the top of strong 
interactions. 

\section{Quark-lepton connections}
%%%%%%%%%%%%%%%%%%%%%%%%%%%%%%%%%%%%%%%%%%%%%%%%%%%%%%%%

\subsection{Quark-lepton symmetry}
%%%%%%%%%%%%%%%%%%%%%%%%%%%%%%%%%%%%%%%%%%%%%%%%%%%%%

There is an apparent correspondence between quarks and leptons. 
Each quark has its own counterpartner in the leptonic sector.  
Leptons can be treated as the 4th color 
following the Pati-Salam $SU(4)_C$ unification symmetry  \cite{pati}.

Further unification  is possible, 
when  quarks and leptons form multiplets of larger gauge group.
The most appealing possibility is SO(10) \cite{so10}, 
where all known components of quarks and leptons
as well as  the RH neutrinos form unique 16-plet.
It is difficult to believe that these features are accidental.
Though, it is not excluded that the
quark-lepton connection has rather  complicated form. 
%{\it e.g.}, of the quark - lepton complementarity \cite{qlc,qlc1}. 

The quark-lepton symmetry is not equivalent to
the quark-lepton unification. Indeed, in the $SU(5)$ GU models
the quark-lepton correspondence
($\nu \leftrightarrow u $, $d \leftrightarrow l$)
is explicitly broken by different $SU(5)$-gauge transformation
properties: $u, u^c \sim {\bf 10}$, 
whereas  $\nu \sim {\bf \bar 5}$, $\nu^c \sim {\bf 1}$, then 
$d \sim {\bf 10}$, $d^c \sim {\bf \bar 5}$ but $l \sim {\bf \bar 5}$,  
$l^c \sim {\bf 10}$. This unification leads to diversity which  is not seen 
in the low energy effective theory. 

The difference of the gauge properties
%(if flavor physics is above the GUT scale) 
can lead to

(i) different mass hierarchies of upper and down quarks,
and also charge leptons and neutrinos~\cite{babu}; 

(ii) different mixings of quarks and leptons. In fact, the 
loopsided mechanism of large mixing realizes this
possibility~\cite{loopside}.\\

Generically, GUT's provide with all ingredients
necessary for the seesaw mechanism:

- RH neutrino components;

- large mass scale;

- lepton number violation.

Besides this, generically GUT's 
give  relations between masses and mixings of leptons and quarks.
They lead to equalities of  masses if a 
single Higgs multiplet is involved in the Yukawa couplings, with 
well  known example being  the $b-\tau$ unification,
$m_b \approx m_\tau$,  at the GUT scale.
%\footnote{Notice the latest determination
%shows some
In general, when several different Higgs
multiplets are involved, one gets ``sum rules''
between masses and mixings of quarks and leptons \cite{sumrule}.

However,   GUT's do not explain the flavor structures.
Apart from some exceptional cases
({\it e.g.}, antisymmetric
representations) no flavor structure is produced by
GUT's. Existing attempts to combine GUT's and
various horizontal or family symmetries (especially
neutrino symmetries) 
have not produced yet substantial results.

\subsection{Quark-lepton universality}
%%%%%%%%%%%%%%%%%%%%%%%%%%%%%%%%%%%%%%%%%%%%%%%%%%%%

Can we speak on the quark-lepton universality in a complete
theory, in spite of big differences of mass and mixing patterns? 
Is it possible that not only the gauge but also 
Yukawa interactions of quark and leptons are very similar?

The idea behind is that the matrix of Yukawa couplings, 
$Y$, has the following form
\be
Y = Y_0 + \delta Y_f, ~~~~ f = u, d, D, l,
\ee
where  $\delta Y_f \ll Y_0$ and $Y_0$ is the universal
matrix for all fermions.
The similarity  (universality)  of quarks and leptons 
is realized in terms of the matrices 
of Yukawa couplings and not of observables - mass ratios and
mixing angles. 
The key point is that similar mass matrices can lead to
substantially different mixing angles and masses (eigenvalues)
if the matrices are nearly singular (rank-1) \cite{sing,dors}. 
The singular matrices are ``unstable''
in a sense that small perturbations can lead to strong variations of
mass ratios and mixing angles (the latter -  in the context of seesaw).

Let us consider the universal structure for the mass matrices
of all quarks and leptons \cite{dors}:
\be
Y_u \sim Y_d \sim Y_D \sim Y_M \sim Y_l \sim Y_0, 
\ee
where  $Y_D$ is the Dirac type neutrino Yukawa matrix,
$Y_M$ is the Majorana type matrix for the RH neutrinos
and  $Y_0$ is the singular matrix. As an important example we take
\be
Y_0 =  
\left(\begin{array}{ccc}
\lambda^4 & \lambda^3 & \lambda^2\\
\lambda^3 & \lambda^2 & \lambda\\
\lambda^2 & \lambda & 1
\end{array}
\right), ~~~~ \lambda \sim 0.2 - 0.3.
\label{anz}
\ee
%Apparently $det Y_0 = 0$ as well as determinants of submatrices are zero.
This matrix has only one non-zero eigenvalue and no physical mixing 
appears at this stage. 

Let us introduce perturbations,  $\epsilon$, in the following form
\be
Y^f_{ij} = Y^0_{ij} (1 + \epsilon_{ij}^f), ~~~ f = u, d, e, \nu, N ,
\label{pert}
\ee
where $Y^0_{ij}$ is the element of the original singular matrix.
This form can be justified, {\it  e.g.}, in context of the Froggatt-Nielsen
mechanism~\cite{fn}. (The key element is the form of perturbations (\ref{pert})
which distinguishes the ansatz (\ref{anz}) from other possible schemes with 
singular matrices.) 
It has been shown that small perturbations
$\epsilon \leq 0.25$ are enough  to explain large difference in mass hierarchies
and mixings of quarks and leptons \cite{dors}.

The seesaw plays crucial role here:
It generates not only small  neutrino masses
but also large lepton mixing. Indeed,
according to the seesaw $m \propto M_R^{-1}$, and  
nearly singular matrix of the RH neutrinos leads
to enhancement of the lepton mixing~\cite{ssenh}.  
%and to flip of sign of mixing
%angle which comes from diagonalization of the neutrino mass matrix.
%As a consequence, the angles from the charged leptons and neutrinos sum up, 
%whereas in quark sector mixing angles from up and down quark mass matrices
%subtract. 

In this approach maximal lepton mixing is accidental.\\

The quark-lepton universality can be introduced differently as 
universality of the {\it mixing matrices} \cite{JS}.  
%(In the lowest order that should be equivalent to certain universality 
%of the mass matrices.) 
One can postulate that in certain ``universality''
basis in the first approximation the mass matrices
of all fermions are diagonalized by the same matrix
$V$ or its charge conjugate $V^*$.

Such a possibility  is inspired  by the $SU(5)$ unification
where leptons and down antiquarks enter the same
5-plet. All the matrices but the matrix for
the charged leptons, $M_l$,  are diagonalized by $V$:
\be
V^{\dagger} M_f V = D_f, ~~~~f = u, d, \nu , 
\label{udn}
\ee
where $D_f$ are  the diagonal mass matrices. For the
charged leptons we have
\be
V^T M_l V^* = D_l.
\label{lll}
\ee
From (\ref{udn}) and (\ref{lll})
one obtains  the SU(5) relation: $M_l = M_d^T$.
(Another version is when
neutrino mass matrix is diagonalized by $V^*$.) 

According to (\ref{udn}, \ref{lll}) in the first approximation one obtains
for the physical mixing matrices
\be
V_{CKM} = V^{\dagger} V = I, ~~~~
V_{PMNS} = V^{T} V.
\label{mixmat}
\ee
The quark mixing is absent, whereas the lepton mixing
is non-trivial and can be large.

In general, the upper and down
fermions are diagonalized by different matrices
$V'$ and $V$. In this case we obtain
\be
V_{CKM} = V^{'\dagger} V, ~~~~
V_{PMNS} = V^{T} V'. 
\label{mixmatgen}
\ee
Now the quark mixing is non-zero in the lowest
order. Furthermore, (\ref{mixmatgen}) leads to
the following relation between mixing matrices:
\be
V_{PMNS} =  V^{T} V V_{CKM}^{\dagger}.
\label{mixcomp}
\ee
So, the quark and lepton mixings are
complementary to $V_{PMNS}^0 = V^{T} V$.
The  matrix  $V_{PMNS}^0$ is symmetric and characterized
by two angles $\phi_1 /2$ and $\phi_2$. It is close to phenomenological
matrix for relatively small values of the angles:  
$\phi_1/2 \sim \phi_2 \sim 20 - 25^{\circ}$.
With the CKM type corrections, as in eq. (\ref{mixcomp}), 
$V_{PMNS}$ gives good description of data and predicts $\sin \theta_{13} > 0.08$
\cite{JS}.

The universal mixing can originate from the  mass matrices
of  particular form which are related to the
universal real matrix $A$:
\be
M_{u,\nu} \approx m D^* A D^*, ~~~
M_{d} \approx m D^* A D, ~~~
M_{l} \approx m D A D^* .
\ee
Here $D \equiv diag(1, i, 1)$.
It happens that the phenomenologically required
structure of the matrix $A$ is very similar to that
in (\ref{anz}).  Such structures can be embedded into
$SU(5)$ and $SO(10)$ models \cite{JS}.

\subsection{Quark-lepton complementarity (QLC)}
%%%%%%%%%%%%%%%%%%%%%%%%%%%%%%%%%%%%%%%%%%%%%%%%%%%%%%%

As it was mentioned in sec. 2.3,  within $1\sigma$  the data are in agreement with 
the 
quark-lepton complementary relations   
\be
\theta_{12} + \theta_C  = \frac{\pi}{4}, ~~~~~  
\theta_{23} + \arcsin V_{cb}  = \frac{\pi}{4},  
%45^{\circ},
\label{qlcrel}
\ee
%The latest determination of the solar mixing angle gives
%$
%\theta_{12} + \theta_C = 46.7^{\circ} \pm 2.4^{\circ}  ~~~(1\sigma)
%$
%which is consistent with maximal mixing angle within $1\sigma$.
%Is the QLC-relation accidental or there is some physics behind, 
%that should include non-trivial quark-lepton connection?

For various reasons it is difficult to expect exact
equalities (\ref{qlcrel}). However certain correlation clearly shows up:

\begin{itemize}

\item
the 2-3 leptonic mixing is close to maximal one because
the 2-3 quark mixing is very small; 

\item
the 1-2 leptonic mixing deviates from maximal one
substantally because the 1-2 quark mixing ({\it i.e.}, Cabibbo
angle) is relatively large.

\end{itemize}

Can it  be accidental?  A general scheme for the QLC relations is that
\be
``{\rm lepton~ mixing} =  {\rm bi-maximal~mixing} - {\rm CKM}'',  
\ee
where the bi-maximal mixing matrix is~\cite{bim}:
\be
U_{bm} = U_{23}^m U_{12}^m =  
\frac{1}{2}
\left(\begin{array}{ccc}
\sqrt{2} & \sqrt{2} & 0\\
-1 & 1 & \sqrt{2}\\
1 & - 1 & \sqrt{2}
\end{array}
\right). 
\label{bimax}
\ee
Here $U_{ij}^m$ is maximal mixing rotation in the $ij$-plane. 
%$U_{bm}$ can play a role of dominant structure
%or matrix in the lowest order.  

Let us consider two possible QLC scenarios which differ by  origin
of the bi-maximal mixing and lead to different predictions.

1). QLC1: The bi-maximal mixing is generated by the neutrino
mass matrix, presumably due to  seesaw. The charged lepton mass matrix
produces  the CKM mixing as a consequence of the q-l symmetry:
$m_l \approx m_d$. Therefore 
\be
U_{PMNS} = U_{CKM}^{\dagger} \Gamma_{\alpha} U_{bm},
\label{qlc1mat}
\ee
where $\Gamma_{\alpha} \equiv diag(1, 1, e^{i\alpha})$ is the  phase 
matrix which appears in general at diagonalization. 
In this case exact relation (\ref{qlcrel}) is not realized since the 
$U_{12}^{CKM}$ rotation matrix should be permuted with $U_{23}^m$ 
in (\ref{qlc1mat}) to reduce (\ref{qlc1mat}) to the standard parametrization form 
(\ref{param}).  
As a consequence, the QLC relation is modified: 
\be
\sin \theta_{12} = \sin (\pi/4 -\theta_C) + 
0.5 \sin \theta_C (\sqrt{2} -1 - V_{cb} \cos \alpha). 
\label{qlc1}
\ee
Numerically (without the RGE effects) we find $\sin^2\theta_{12} = 0.3345$ 
for $\alpha \sim 90^{\circ}$ and   $\sin^2\theta_{12} = 0.330$ for  $\alpha = 0$.
This is practically indistinguishable from the tri-bimaximal mixing 
prediction  $\sin^2\theta_{12} = 0.3333$.

Let us stress that practically the  same predictions for 1-2 mixing are obtained 
from 
two different combinations of matrices:  
\be
U_{23}^m U_{12}(\arcsin(1/\sqrt{3}))~~~~ {\rm and} ~~~~U_{12}(\theta_C)U_{23}^m 
U_{12}^m    
\ee
which are completely independent. 
Therefore an equality of the predictions is 
just accidental coincidence.  This means that one of the two approaches 
(QLC1 or tri-bimaximal mixing) is wrong. 
To some extend that can be tested by measuring the 1-3 mixing.  
In the QLC1-scenario one obtains
\be
\sin^2 \theta_{13} = 0.5 \sin^2 \theta_C \approx 0.0245,  
\ee
whereas the tri-bimaximal mixing implies  $\sin^2 \theta_{13} = 0$
unless some corrections are introduced.\\

2). QLC2: Maximal mixing comes from the charged lepton mass matrix
and the CKM mixing originates from the neutrino mass matrix due to
the q-l symmetry: $m_D \sim m_u$ (assuming also that in the context of seesaw
the RH neutrino mass matrix does not influence 
mixing). Consequently, 
\be
U_{PMNS} = U_{bm} \Gamma_{\alpha} U_{CKM}^{\dagger}. 
\ee
In this case the QLC relation for 1-2 mixing is satisfied precisely:
$\sin \theta_{12} = \sin (\pi/4 -\theta_C)$. 
Now $\sin^2 \theta_{13}  \approx  \sin^2 \theta_{12} V_{cb}^2$ is extremely 
small.

All three predictions for 1-2 mixing (from QLC1, QLC2 and tri-bimaximal mixing) 
are within $1\sigma$ errors from the b.f. point. The tri-bimaximal mixing 
and  QLC1 predictions almost coincide,  the b.f. value is in between
the QLC2 and two other predictions: $\theta_{12}(QLC2) < \theta_{12}^{exp} < \theta_{12}(QLC1) \approx \theta_{12}(tbm)$.
To disentangle these two possibilities
one needs to measure the 1-2 mixing with accuracy
$\Delta \theta_{12} \sim  1^{\circ}$ or
$\Delta \sin^2 \theta_{12} \sim  0.015$ ($5\%$).\\

%%%%%%%%%%%%%%%%%

There are two main issues related to the QLC relations:

(1) origin of the bi-maximal mixing; 

(2) mechanism of propagation  of the CKM mixing 
from the quark sector  to the  lepton one.
The problem here is big  difference of mass ratios
of the quarks and leptons:  $m_e/m_\mu = 0.0047$, 
$m_d/m_s = 0.04 - 0.06$,  as well as difference of masses of muon and
s-quark at the GU scale. 
%So,  difference of the mass eigenvalues should be
%reconciled with equal (close) mixings. 
This means that mixing should weakly depend on or be independent of masses. 

So, if not accidental,  the QLC relation may have the 
following implications:

- the  quark-lepton symmetry, 

- existence of some additional
structure which produces the bi-maximal mixing, 

- mass matrices with weak correlation of the mixing angles and  mass eigenvalues. 

Alternatively, it may imply certain flavor physics with
$\sin \theta_C$ being  the ``quantum'' of this physics.

In majority of  models  proposed so far,  the approximate 
QLC relation appears as a result of interplay of different independent 
factors or as sum of several independent contributions. 
From this point of view the QLC relation is accidental.

\section{Effects of new neutrino states}
%%%%%%%%%%%%%%%%%%%%%%%%%%%%%%%%%%%%%%%%%%%%%%%%%

Effects of new neutrino states (singlets of the SM symmetry group) 
depend on their masses.
Superheavy new states essentially decouple.
%An important  criteria is that $M_S \gg Q$,
%where $Q$ is typical energy release
%in the neutrino processes. Here we can take $Q \sim m_W$ -
%the mass of the $W$-boson. (One consider also  production
%of even heavier neutrinos  in  the high energy collisions.)
These states are not produced in laboratory experiments,
but they  can lead to  indirect effects:

- modify substantially the
mass matrix of active neutrinos; 

- violate  universality of
the weak interactions, 
{\it etc.}.

For relatively small masses,   
say $M_S \ll  m_W$, 
these new states can be produced in reactions
thus leading to direct effects but also
they modify the mass matrix of active neutrinos.
Light new states with $m_S \sim m_{\nu}$ can lead to  non-trivial 
oscillation effects.

Here we consider two applications of
possible existence of new neutrinos states. They  realize 
an idea that these states play the role of 
additional structures which lead to
substantial difference of quark
and lepton properties.

\subsection{Screening of Dirac structure}
%%%%%%%%%%%%%%%%%%%%%%%%%%%%%%%%%%%%%%%%%%%%%%%%%%%%%%%%%%%%%%%%%%%%%

%The quark -lepton symmetry manifests  as certain relation
%(similarity) between the Dirac mass matrices of quarks and leptons,
%and it is this  feature which creates problem for explanation of
%strongly different mixings and possible existence of the ``neutrino'' symmetries.
%Let us  consider an extreme case when in spite of the q-l unification, 
%the Dirac structure in the lepton sector is completely eliminated -  
%``screened'' \cite{scre}.

Let us introduce one heavy neutral  state $S$ for each generation and
consider mass matrix in the basis $(\nu, N^c,  S)$ of the following form 
\be
m = 
\left(\begin{array}{ccc}
0 & m_D & 0\\
m_D^T & 0 & M_D^T\\
0 & M_D & M_S
\end{array}
\right). 
\label{dss}
\ee
Here $M_S$ is the Majorana mass matrix of new fermions. 
Such a structure can be formed by a 
lepton number violated in the $M_S$ and some additional 
symmetry which forbids also 13-element. 

For $m_D \ll M_D \ll M_S$ the matrix  leads to the double (cascade)
seesaw mechanism~\cite{dss}:
\be
m_{\nu} = m_D^T M_D^{-1 T} M_S M_D^{-1 } m_D, 
\label{doubless}
\ee
and the mass matrix of RH neutrinos becomes $M_R = - M_D M_S^{-1} M_D^{T}$. 
If two Dirac mass matrices are proportional each other,  
\be
M_D = A^{-1} m_D, ~~~~ A \equiv  v_{EW}/V_{GU}, 
\label{propo}
\ee
they cancel in (\ref{doubless}) and we obtain
\be
m_{\nu} = A^2 M_S.
\ee
That is, the structure of light neutrino mass matrix is determined by
$M_S$ immediately and does not depend on the Dirac mass matrix (the later is 
screened).
The seesaw mechanism provides  scale of
neutrino masses but not
the flavor structure of the mass matrix.

Notice that screening does not depend on the scale of $M_S$ and in fact 
$M_S \ll M_D$ is also possible.
However it is natural to  assume that $M_D$ is at the GUT scale, and 
$M_S$ is at the Planck scale $M_{Pl}$ 
which leads to correct values of the light neutrino masses. 
It can be shown that at least in SUSY version the  radiative corrections do 
not destroy screening \cite{scre}. The relation (\ref{propo}) can be a consequence 
of Grand Unification with extended gauge group or/and certain 
flavor symmetry~\cite{scre,kim}.

Structure of the light neutrino mass matrix depends now on $M_S$ which can be 
related to some physics at the  Planck scale, and consequently, lead to ``unusual'' 
properties of neutrinos. In particular,

(i) certain symmetry of $M_S$ can be the origin of ``neutrino'' symmetry;

(ii) the matrix  $M_S \propto I$  leads to the quasi-degenerate 
mass spectrum;

(iii) $M_S$ can be the origin of bi-maximal mixing 
thus leading to the QLC relations,  
if the charged lepton mass matrix generates the CKM rotation.

%It allows to reconcile the q-l symmetry with
%strong difference of mixings of leptons and quarks. 

\subsection{New states and induced mass matrix}
%%%%%%%%%%%%%%%%%%%%%%%%%%%%%%%%%%%%%%%%%%%%%%

Suppose the active neutrinos acquire ({\it e.g.}, via seesaw)
the Majorana mass matrix $m_a$. Consider one sterile neutrino, 
$S$,   with Majorana mass $m_S$
and mixing  with active neutrinos characterized by ``vector'' of masses 
$\bar{m}_{S} \equiv ( m_{eS}, m_{\mu S}, m_{\tau S})$. 
Essentially in the basis $(\nu, N^c,  S)$ 
this corresponds to the mass matrix of the form 
\be
m =
\left(\begin{array}{ccc}
0     & m_D & \bar{m}_S \\
m_D^T & M_R & 0 \\
\bar{m}_S  & 0   & m_S
\end{array}
\right).
\label{indm}
\ee

If $m_S \gg m_{iS}$,  
then after decoupling of $S$ the mass matrix of active neutrinos becomes 
\be
m_{\nu} = m_a + m_I,  
\ee
where the last term is the matrix induced by $S$: 
\be
m_I = \frac{1}{m_S} \bar{m}_{S}^T  \bar{m}_{S}.   
\label{induce}
\ee
The induced matrix has zero determinant 
and therefore can be an origin of  singular structures. 

Introducing  the active-sterile mixing angle
$\theta_S$  as  
\be
\sin\theta_S = \bar{m}_S/m_S, 
\label{stmix}
\ee
we can rewrite the elements of induced matrix as 
\be
m_I  \sim \sin^2 \theta_S m_S.
\label{indmix}
\ee

The induced matrix may turn out to be   the ``missed'' element which leads to 
the difference 
of mixings of quarks and leptons. Let us consider several possibilities. 

1). Suppose $\bar{m}_{S} \propto (0, 1, 1)$, then the induced matrix 
reproduces the dominant block of the active neutrino mass matrix 
for the normal mass hierarchy: 
\be
m_\nu =
\frac{\sqrt{\Delta m^2_{32}}}{2}
\left(\begin{array}{ccc}
... &  ...  & ... \\
...  & 1 & 1\\
...  &  1 & 1
\end{array}
\right),
\label{domeff}
\ee
where ``dots''  denote small parameters. 
In this case one can realize a possibility 
that the original active neutrino mass matrix, $m_a$, has hierarchical structure 
with small mixings being similar to the quark mass matrices. 
From eqs. (\ref{domeff}) and (\ref{indmix}) we find 
\be
\sin^2 \theta_S m_S = \frac{1}{2} \sqrt{\Delta m^2_{32}} \sim 0.025~ {\rm eV}.
\label{domb}
\ee

2).  Let us assume that couplings of $S$ with active neutrinos are universal - 
flavor ``blind'':
\be
\bar{m}_{S} \propto (1, 1, 1). 
\ee
Then the induced matrix has form: $m_{I} \propto  D$, 
where $D$ is the democratic matrix - the second 
matrix in (\ref{tribi}). 
Suppose that the original active neutrino mass matrix has structure 
of the first matrix in (\ref{tribi}). 
Then the sum,  $m_{\nu} = m_a + m_{I}$,  reproduces the mass
matrix for the tri-bimaximal mixing (\ref{tribi}).
In this case, according to (\ref{tribi}),  
the parameters of $S$  should satisfy relation  
\be
\sin^2 \theta_S m_S = \frac{1}{3} \sqrt{\Delta m^2_{21}} \sim 0.003~ {\rm eV}.
\label{subdomb}
\ee

With two sterile neutrinos whole structure (\ref{tribi})
can be obtained.

3). New neutrino states are irrelevant if
$m_{iS}m_{jS}/m_S \ll (m_a)_{ij}$ 
or
\be
\sin^2 \theta_S~ m_S < ~ 0.001~ {\rm eV}. 
\label{smixing}
\ee

Clearly, the presence of induced contribution changes 
implications of the neutrino results~\cite{abdel,renata}.  Since $S$ is beyond the 
SM structure 
extended by RH neutrinos,  it may be easier to realize ``neutrino'' symmetries as 
a consequence of certain symmetry of $S$ couplings with active neutrinos.

\begin{figure}[htb]
\epsfxsize=10cm
\begin{center}
\leavevmode
\epsffile{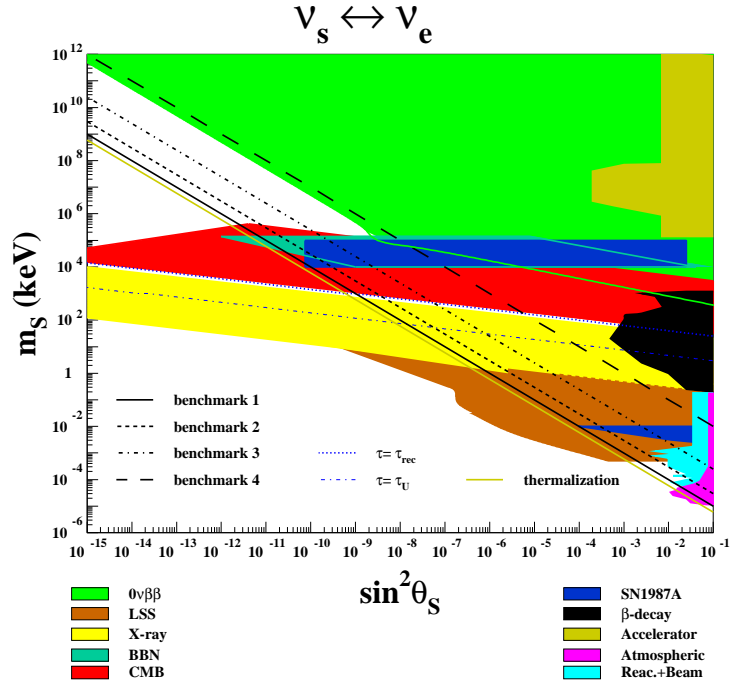}
\end{center}
\vglue -3cm
\caption{ The benchmark lines of induced masses given in eqs. 
(\ref{domb}), (\ref{subdomb}), (\ref{smixing}) 
  versus the current astrophysical, cosmological and
  laboratory bounds on $\nu_S -\nu_e$ mixing.
  The colored regions are excluded.  The
  ``thermalization'' line and the two decay lines $\tau_S = \tau_{\rm
    rec}$ and $\tau_S = \tau_U$ are also shown.}
\label{fig1}
\end{figure}

\begin{figure}[htb]
\epsfxsize=10cm
\begin{center}
\leavevmode
\epsffile{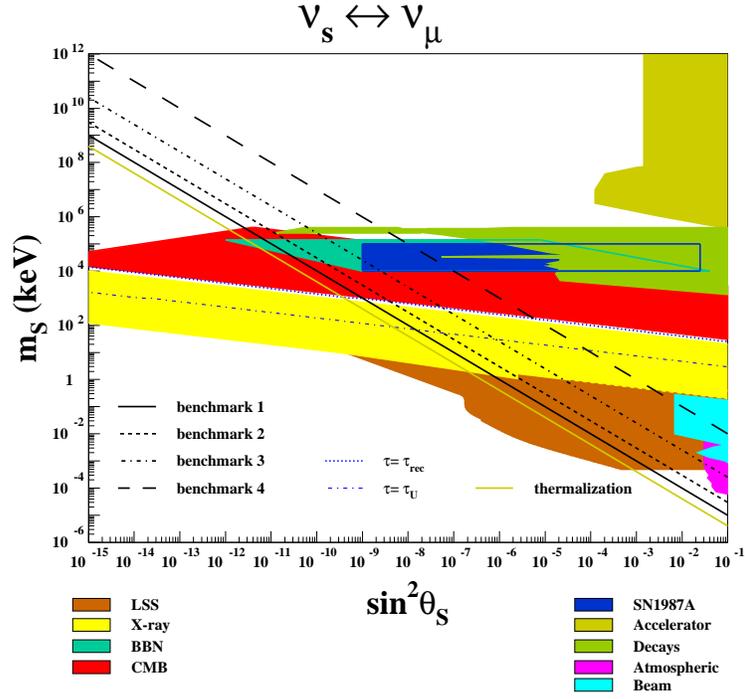}
\end{center}
\vglue -0.5cm
\caption{The same as in Fig.\ref{fig1} but for $\nu_S -\nu_\mu$ mixing.}
\label{fig2}
\end{figure}

In figs. \ref{fig1} and \ref{fig2} we show lines of constant induced masses 
in the plane $\sin^2 \theta_S -  m_S$  which 
are given by the conditions (\ref{domb}), (\ref{subdomb}), (\ref{smixing})
as well as the line $\sin^2 \theta_S m_S < ~ 0.5$ eV which coresponds 
to maximal allowed value of the matrix elements. 
We confront these lines with various cosmological, astrophysical and laboratory 
bounds on the parameters of new neutrino states (see ref.\cite{renata} for details).

%%%%%%%%%%ffff1%%%%%%%%%%%%%%%%%%%%%%%%%%%%%%%%%%%%%%%%%%%%%%%%%%%%%%%%%%%
\vglue 1.8cm
%%%%%%%%%%%%%%%%%%%%%%%%%%%%%%%%%%%%%%%%%%%%%%%%%%%%%%%%%%%%%%%%%%%

According to figs. \ref{fig1} and \ref{fig2} two regions are allowed:

1). Small masses window:  $m_S \sim (0.5 - 1)$ eV 
and $\sin^2 \theta_S = 0.001 - 0.1$,  
where direct and indirect effects are comparable. 
This window  is disfavored by results 
of recent analysis of cosmological data \cite{Uros06}, 
and it is closed if the Big Bang nucleosynthesis bound on the effective 
number of neutrino species 
$N_{\nu} < 4$ is taken.   

Notice that there are various ways to avoid the cosmological bounds which however 
imply an existence of additional physics beyond the Standard model \cite{renata}.

%%%%%%%%%%%%%%%ffff2%%%%%%%%%%%%%%%%%%%%%%%%%%%%%%%%%%%%%%%%%%%%%%%%%%%%%%
%\vglue -1cm
%%%%%%%%%%%%%%%%%%%%%%%%%%%%%%%%%%%%%%%%%%%%%%%%%%%%%%%%%%%%%%%%%%%

2). Large masses range: $m_S > 300$ MeV and $\sin^2 \theta_S < 10^{-9}$.   
Here direct mixing effects are negligible and the presence of 
new states can not be verified.

\section{Summary}
%%%%%%%%%%%%%%%%%%%%%%%%%%%%%%%%%%%%%%%%%%%%%%%%%%

Comparison of the  properties of the quarks and leptons
shows similar gauge characteristics  and  
strong difference of mass and mixing patterns.  

\newpage

There are several observations which 
(if not accidental) can strongly influence implications of the results.  
Those include possible presence of special leptonic (neutrino)
symmetries; particular (tri-bimaximal)
form of neutrino mixing matrix; 
quark-lepton complementarity relations. 
These features may indicate that quarks and leptons are
fundamentally different
and some new structures of theory exist beyond the seesaw.

Mixing with new neutrino states can play the role
of this additional structure. In particular, it can 

- produce screening of the Dirac structure; 

- generate the  induced matrix of active neutrinos with
certain symmetry properties. The induced matrix can lead to 
enhancement of lepton mixings, to generation of the dominant 
block of the mass matrix  in the case of normal mass hierarchy,  
or to various subdominant structures, {\it e.g.}, 
for the tri-bimaximal mixing.

Still the approximate quark-lepton universality can be
realized. In this case, the dominant mass or mixing matrices are 
the same for all fermions and small (of the order $\sin \theta_C$) corrections 
can produce whole difference. The seesaw mechanism plays the key role in getting of 
large lepton mixing.

\section{Acknowledgements}

This work has been supported in part by the Alexander von Humboldt Foundation 
(the Humboldt research award).


\begin{thebibliography}{99}

\bibitem{bari}G. L. Fogli et al,  hep-ph/0506083.

\bibitem{sno}SNO Collaboration (B. Aharmim et al.). {\it Phys. Rev.} C 
{\bf 72}, 055502 (2005). 

\bibitem{sv}A. Strumia, F. Vissani, {\it Nucl. Phys.} B {\bf 726}, 294 (2005). 

\bibitem{atm} Super-Kamiokande Collaboration (Y. Ashie et al.), 
{\it Phys. Rev.} D {\bf 71} 112005, (2005)

\bibitem{orl}O. L. G. Peres, A. Yu. Smirnov,  {\it Phys. Lett.} B {\bf 456}, 204 
(1999);  {\it Nucl. Phys.} B {\bf 680}, 479 (2004). 

\bibitem{concha}M. C. Gonzalez-Garcia, M. Maltoni, A. Yu. Smirnov, 
{\it Phys. Rev.} D {\bf 70}, 093005 (2004). 

%\bibitem{cos}U. Seljak et al.. {\it Phys. Rev.} D {\bf 71}, 103515 (2005). 

\bibitem{Uros06}
  U.~Seljak, A.~Slosar and P.~McDonald,
  %``Cosmological parameters from combining the Lyman-alpha forest with CMB,
  %galaxy clustering and SN constraints,''
  arXiv:astro-ph/0604335. 

\bibitem{hm}H.V. Klapdor-Kleingrothaus, et al, {\it Phys. Lett.} B 
{\bf 586}, 198 (2004). 


\bibitem{tbm}
L. Wolfenstein,  {\it Phys. Rev.} D {\bf 18}, 958 (1978); 
P. F. Harrison, D. H. Perkins and W. G. Scott, 
{\it Phys. Lett.} B {\bf 458}, 79 (1999), 
{\it Phys. Lett.} B {\bf 530}, 167 (2002).  

\bibitem{model}E. Ma,
{\it Mod. Phys. Lett.} A {\bf 17}, 2361 (2002); 
E. Ma, G. Rajasekaran, {\it  Phys. Rev.} D {\bf 64} 113012, (2001);
K.S. Babu, E. Ma, J.W.F. Valle,
{\it Phys. Lett.} B {\bf 552}, 207 (2003);
%%\bibitem{model} For some recent publications see:  
 W.~Grimus and L.~Lavoura, {\it JHEP} {\bf 0508}, 013 (2005); 
K.~S.~Babu and X.~G.~He, hep-ph/0507217; 
E.~Ma, {\it Mod.\ Phys.\ Lett.} A {\bf 20}, 2601 (2005). 
G. Altarelli and F. Feruglio, hep-ph/0507217; 
H. G. He, Yong-Yeon Keum and R. R. Volkas, hep-ph/0601001. 

\bibitem{qlc}A. Yu. Smirnov, hep-ph/0402264; 
M. Raidal, {\it Phys. Rev. Lett.}  {\bf 93}, 161801 (2004).   

\bibitem{qlc1}
H. Minakata, A. Yu. Smirnov, {\it Phys. Rev.} D {\bf 70}, 
073009 (2004). 

\bibitem{mutau}
T.~Fukuyama and H.~Nishiura, hep-ph/9702253;   R.~N.~Mohapatra and 
S.~Nussinov, {\it Phys.\ Rev.}\ D {\bf 60}, 013002 (1999); 
E.~Ma and M.~Raidal, {\it Phys.\ Rev.\ Lett.}\  {\bf 87}, 011802 (2001);  
C.~S.~Lam, {\it Phys.\ Lett.}\ B {\bf 507}, 214 (2001). 

\bibitem{anj05}
  A.~S.~Joshipura,
  %``Universal 2-3 symmetry,''
  arXiv:hep-ph/0512252.

\bibitem{renata}
  A.~Y.~Smirnov and R.~Zukanovich Funchal,
  %``Sterile neutrinos: Direct mixing effects versus induced mass matrix of
  %active neutrinos,''
  arXiv:hep-ph/0603009.


\bibitem{sees} P. Minkowski, {\it Phys. Lett.} B {\bf 67} 421 (1977); 
T. Yanagida, in {\it Proc. of Workshop on Unified Theory and Baryon
number in the Universe}, eds. O. Sawada and A. Sugamoto, KEK, Tsukuba, (1979);
M. Gell-Mann, P. Ramond and R. Slansky,  in {\it Supergravity}, eds P. 
van Niewenhuizen and
D. Z. Freedman (North Holland, Amsterdam 1980);
P. Ramond, {\it  Sanibel talk}, retroprinted as hep-ph/9809459;
S. L. Glashow, in {\it Quarks and Leptons}, Carg\`ese lectures, eds M. L\'evy,
(Plenum, 1980, New York) p. 707;
R. N. Mohapatra and G. Senjanovi\'c, {\it Phys. Rev. Lett.} {\bf 44}, 912 (1980).


\bibitem{pati}J. C. Pati and A. Salam, {\it Phys. Rev.} D {\bf 10}, 275 (1974). 

\bibitem{so10}H. Georgi, {\it In Coral Gables 1979 Proceeding, Theory and experiment 
in high energy physics}, New York 1975, 329 and H. Fritzsch and P. Minkowski, Annals 
Phys. {\bf 93} 193 (1975). 


\bibitem{babu}K. S. Babu and S. M. Barr, Phys. Lett. B {\bf 381} (1996) 202.  

\bibitem{loopside}C. H. Albright, K. S. Babu and S. M. Barr, Phys. Rev. Lett. 
{\bf 81} (1998) 1167. 

\bibitem{sumrule}B. Bajc, G. Senjanovic and F. Vissani, Phys. Rev. Lett. {\bf 90} 
(2003) 051802.   

\bibitem{sing}E.~K.~Akhmedov, et al.,
 {\it  Phys.\ Lett.}\ B {\bf 498}, 237 (2001); R. Dermisek,  
{\it Phys. Rev.} D {\bf 70}, 033007 (2004). 

\bibitem{dors}I. Dorsner, A.Yu. Smirnov,  
{\it Nucl. Phys.} B {\bf 698}, 386 (2004). 


\bibitem{fn}C. D. Froggatt and H. B. Nielsen, 
{\it Nucl. Phys.} B {\bf 147}, 277 (1979).


\bibitem{ssenh}A. Yu. Smirnov, {\it Phys. Rev.} D {\bf 48}, 3264 (1993).

\bibitem{JS}
  A.~S.~Joshipura and A.~Y.~Smirnov,
  %``Quark-Lepton universality and large leptonic mixing,''
  arXiv:hep-ph/0512024.

\bibitem{bim}F. Vissani, hep-ph/9708483; 
V.~D.~Barger, et al, {\it  Phys.\ Lett.}\ B {\bf 437}, 107 (1998).

\bibitem{dss}R. N. Mohapatra, {\it Phys. Rev. Lett.}  {\bf 56}, 561 (1986); 
R. N. Mohapatra and J. W. F. Valle, {\it Phys. Rev.} D {\bf 34}, 1642 (1986).

\bibitem{scre}M. Lindner, M. A. Schmidt, A. Yu. Smirnov, {\it JHEP} {\bf 0507}, 
048 (2005). 

\bibitem{kim}O. Vives, hep-ph/0504079; J. E. Kim and J. C. Park, hep-ph/0512130. 

\bibitem{abdel}K.R.S. Balaji, A. Perez-Lorenzana, A.Yu. Smirnov,  
{\it Phys. Lett.} B {\bf 509}, 111 (2001). 


\end{thebibliography}
\end{document}